\documentclass[aps,prd,preprintnumbers,nofootinbib,twocolumn]{revtex4}

\usepackage{amsmath}
\usepackage{amsfonts}
\usepackage{amssymb}
\usepackage{color}

\begin{document}

\title{Minimal Length in quantum gravity and gravitational measurements}
\author{Ahmed Farag Ali$^{1,2}$}\email[]{ahmed.ali@fsc.bu.edu.eg, afali@fsu.edu}
\author{Mohammed M. Khalil$^{3}$} \email[]{moh.m.khalil@gmail.com}
\author{Elias C. Vagenas$^{4}$}\email[]{elias.vagenas@ku.edu.kw}
\affiliation{$^1$Department of Physics, Florida State University, Tallahassee, FL 32306,  USA.
\\ $^2$Department of Physics, Faculty of Science,\\Benha University, Benha, 13518, Egypt\\\vspace{-0.2ex}}
\affiliation{$^3$Department of Electrical Engineering,\\ Alexandria University,  Alexandria 12544, Egypt\\}
\affiliation{$^4$Theoretical Physics Group, Department of Physics, Kuwait University, P.O. Box 5969, Safat 13060, Kuwait}

\begin{abstract}
\par\noindent
The existence of a minimal length is a common prediction of
various theories of quantum gravity. This minimal length
leads to a modification of the Heisenberg uncertainty principle
to a Generalized Uncertainty Principle (GUP). Various studies showed that
a GUP modifies the Hawking radiation of black holes.
In this paper, we propose a modification of the Schwarzschild
metric based on the modified Hawking temperature derived from the GUP.
Based on this modified metric, we calculate corrections to the
deflection of light, time delay of light, perihelion precession,
and gravitational redshift. We compare our results with gravitational
measurements to set an upper bound on the GUP parameter.
\end{abstract}

\maketitle

\section{Introduction}
\vspace{-2ex}
\par\noindent
Various approaches to quantum gravity (QG) are expected to play a crucial
role in revealing some characteristic features of the fundamental quantum theory of gravity.
One common feature among most of these approaches, such as string theory and black hole physics
\cite{Amati:1988tn,Scardigli:1999jh,Garay:1994en,Kempf:1994su,Kempf:1996fz,Brau:1999uv,Maggiore:1993rv},
is the existence of a minimal observable length, i.e. Planck length $l_p$. The existence of a
minimal length leads to the modification of the Heisenberg uncertainty 
principle to a Generalized Uncertainty Principle (GUP)
which includes an additional quadratic term in momentum as follows \cite{Amati:1988tn,Scardigli:1999jh,Garay:1994en,Kempf:1994su,Kempf:1996fz,Brau:1999uv,Maggiore:1993rv}
\begin{equation}
\label{gupquadratic}
\Delta x\Delta p \geq \frac{\hbar}{2}(1+\beta\Delta p^2)
\end{equation}
where $\beta=\beta_0 l_p^2/\hbar^2$, $\beta_0$ is a dimensionless constant, and $l_p=1.6162\times 10^{-35}\text{m}$ is the Planck length.
Phenomenological aspects of GUP effects have been analyzed in several contexts such as the
self-complete character of gravity {\cite{Dvali:2010bf,Isi:2013cxa}, the conjectured black hole productions at the terascale \cite{Mureika:2011hg,Nicolini:2011nz} and the modifications of
neutrino oscillations \cite{Sprenger:2010dg}.

Recently, another interesting form of the GUP has been proposed in \cite{Ali:2009zq,Das:2010zf,Ali:2011fa} to be
consistent with doubly special relativity, string theory, and black hole physics.
This suggested form of GUP includes a linear term in momentum, and leads to a maximum
observable momentum in addition to a minimal length
\begin{eqnarray}
\label{gup}
[x_i, p_j] = i \hbar\hspace{-0.5ex} \left[  \delta_{ij}\hspace{-0.5ex}
- \hspace{-0.5ex} \alpha\hspace{-0.5ex}  \left( p \delta_{ij} +
\frac{p_i p_j}{p} \right)
+ \alpha^2 \hspace{-0.5ex}
\left( p^2 \delta_{ij}  + 3 p_{i} p_{j} \right) \hspace{-0.5ex} \right]~
\end{eqnarray}
where $\alpha=\alpha_0 l_p/\hbar$, and $\alpha_0$ is a dimensionless constant.
The upper bounds on the parameter $\alpha_0$ have been calculated in \cite{Ali:2011fa}
and it was proposed that GUP may introduce an intermediate length scale between
Planck scale and electroweak scale. Recent proposals suggested that these bounds
can be measured using quantum optics techniques in \cite{Pikovski:2011zk} and using gravitational wave techniques \cite{Marin:2014wja,Marin:2013pga} which may be
considered as a milestone in quantum gravity phenomenology. In a series of papers, various phenomenological implications of the new model of GUP were investigated
\cite{Majhi:2013koa,Amelino-Camelia:2013fxa,Majumder:afa,Nozari:2012gd,Ching:2012vq,Ali:2013qza}.
A detailed review along the mentioned lines of minimal length theories and quantum gravity phenomenology can be found in \cite{Hossenfelder:2012jw,Sprenger:2012uc,AmelinoCamelia:2008qg}.

Very recently, Scardigli and Casadio \cite{Scardigli:2014qka} proposed a modification of the Schwarzschild metric
to reproduce the modified Hawking temperature \cite{Adler:2001vs,Bolen:2004sq,Medved:2004yu,Myung:2006qr,Nouicer:2007jg} which was derived from the GUP
of Eq. \eqref{gupquadratic}. This modification of the metric takes the form
\begin{equation}
\label{metric}
d\tau^2=F(r)dt^2-\frac{1}{F(r)}dr^2-r^2d\Omega^2,
\end{equation}
with
\begin{equation}
\label{frSC}
F(r)=1-\frac{2GM}{r}+\epsilon\frac{G^2M^2}{r^2},
\end{equation}
and from comparing the Hawking temperature derived from the GUP with the one 
derived from the modified metric they concluded that $\beta=-\pi^2\epsilon^2M^2/4M_p^2$, 
where $M_p$ is the Planck mass. There is a problem in the form of $F(r)$ in Eq. \eqref{frSC}, which implies that the horizon is at a different value from $r_s=2GM$ contrary to many arguments based on the GUP \cite{Maggiore:1993rv,Scardigli:1999jh,AmelinoCamelia:2005ik}. More importantly, it leads to a negative GUP parameter $\beta<0$ {\footnote{The negativity of the GUP parameter $\beta$ was first encountered in the study of the uncertainty principle given by Eq. \eqref{gupquadratic} when formulated on a crystal lattice \cite{Jizba:2009qf}.}, which is inconsistent with almost all the motivations that led to the GUP \cite{Amati:1988tn,Scardigli:1999jh,Garay:1994en,Maggiore:1993rv,Kempf:1994su,Kempf:1996fz,Brau:1999uv}; the GUP implies a minimal length of $\Delta x\geq\hbar\sqrt{\beta}$ which would be imaginary if $\beta$ is negative.

In this paper, we continue our investigations of the phenomenological implications of GUP that was studied in \cite{Ali:2011fa}. We propose a modification to the Schwarzschild metric to reproduce the modified Hawking temperature derived from the GUP in Eq. \eqref{gup}. We use a more general form for $F(r)$ than the one used by Scardigli and Cassadio (see Eq.  \eqref{assumedFr} below). We do not assume a modification with $1/r^2$ dependence. Instead, we consider a metric with a general $1/r^n$ dependence. In addition, this form leads to a horizon at the usual value $r_s=2GM$, and yields a positive GUP parameter.

In the following sections, we review the derivation of the modified Hawking temperature from the GUP, and find the relation between the GUP parameter $\alpha$ and the metric. Then, we use this metric to find corrections to the general relativistic results of the deflection of light, time delay of light, perihelion precession, and gravitational redshift. We compare our results with experiment to set upper bounds on the GUP parameter $\alpha_0$.

\section{Hawking Temperature from GUP}
\vspace{-2ex}
\par\noindent
The Hawking temperature of a black hole takes the well-known form \cite{Hawking:1974sw}
\begin{equation}
\label{tempH}
T_H=\frac{1}{8\pi GM},
\end{equation}
where, from now on, we use natural units, in which $c=1$, $\hbar=1$, $G=6.708\times10^{-39}\text{GeV}^{-2}$ and $l_p=\sqrt{G}=8.19\times10^{-20}\text{GeV}^{-1}$.

\par\noindent
The GUP modifies the Hawking temperature; that modification was derived in several papers for different forms of the GUP \cite{Adler:2001vs,Bolen:2004sq,Medved:2004yu,Myung:2006qr,Nouicer:2007jg,Majumder:2011xg,Ali:2012hp,Ali:2012mt}. We will follow the derivation in \cite{Majumder:2011xg,Ali:2012mt,Ali:2012hp}.

\par\noindent
We start by rewriting Eq. \eqref{gup} as \cite{Majumder:2011xg}
\begin{equation}
\label{gupsimple}
\Delta x \Delta p \geq \frac{1}{2}\left[1-\alpha \Delta p + 4\alpha^2 (\Delta p)^2\right].
\end{equation}
Solving Eq. \eqref{gupsimple} for $\Delta p$ and expanding to second order in $\alpha$, we get a momentum uncertainty of
\begin{equation}
\label{deltap}
\Delta p\geq \frac{1}{2\Delta x}-\alpha \frac{1}{4\Delta x^2} +\alpha^2 \frac{5}{8\Delta x^3}.
\end{equation}
\par\noindent
According to \cite{Medved:2004yu,Adler:2001vs,Cavaglia:2003qk}, a photon is used to ascertain the position of a quantum particle of energy $E$
and according to the argument in \cite{AmelinoCamelia:2004xx} which demonstrated that the uncertainty principle
$\Delta p \geq 1/ \Delta x$  can be written as a lower bound $E\geq 1/ \Delta x$.
The uncertainty principle $\Delta p\Delta x\geq 1/2$ leads to an energy uncertainty of $\Delta E\geq 1/2\Delta x$. Similarly, from Eq. \eqref{deltap} we get
\begin{equation}
\Delta E\geq \frac{1}{2\Delta x}-\alpha \frac{1}{4\Delta x^2} +\alpha^2 \frac{5}{8\Delta x^3}.
\end{equation}
\par\noindent
The energy uncertainty can be viewed as the energy of the emitted photon from the black hole, and thus as its characteristic temperature $T=\Delta E$. Taking the uncertainty in position to be proportional to the Schwarzschild radius $\Delta x=\mu r_s=2\mu GM$ gives the temperature
\begin{equation}
T\simeq \frac{1}{4\mu GM}-\alpha \frac{1}{16\mu^2G^2M^2} +\alpha^2 \frac{5}{64\mu^3G^3M^3}.
\end{equation}
Comparing this equation with the standard Hawking temperature \eqref{tempH}, we see that $\mu =2\pi$ and the temperature from the GUP is
\begin{equation}
\label{tempgup}
T\simeq\frac{1}{8\pi GM}\left(1-\frac{\alpha}{8\pi GM}+5\left(\frac{\alpha}{8\pi GM}\right)^2 \right).
%
%
\end{equation}

\section{Modified Schwarzschild Metric}
\vspace{-2ex}
\par\noindent
The standard Hawking temperature can be derived from the metric using the surface gravity $\kappa$
\begin{equation}
T_H=\frac{\kappa}{2\pi}
\end{equation}
\par\noindent
where $\kappa$ is related to the metric by (\cite{gron2007einstein}, p. 246)
\begin{equation}
\kappa=\lim_{r\to r_s}\sqrt{-\frac{1}{4}g^{rr}g^{tt}g_{tt,r}^2}.
\end{equation}
For the Schwarzschild metric, Eq. \eqref{metric}, the surface gravity is simply half the derivative of $F(r)$ at the Schwarzschild radius
\begin{equation}
\kappa=\frac{1}{2}F'(r_s),
\end{equation}
Using $F(r)=1-2GM/r$, we get the standard Hawking temperature
\begin{equation}
T_H=\frac{1}{4\pi} F'(r_s)=\frac{1}{8\pi GM}.
\end{equation}
\par\noindent
We can follow the same argument backwards; start from the modified temperature \eqref{tempgup} and look for the metric that reproduces it. We assume the metric takes the same form as Eq. \eqref{metric} but $F(r)$ is modified to
\begin{equation}
\label{assumedFr}
F(r)=\left(1-\frac{2 GM}{r}\right) \left(1+\eta  \left(\frac{2 GM}{r}\right)^n\right),
\end{equation}
where $\eta$ is a constant $\ll1$, and $n$ is an integer $\geq 0$. Differentiating $F(r)$ at $r_s=2GM$ we get the temperature
\begin{equation}
T=\frac{1}{4\pi}F'(r_s)=\frac{1+\eta}{8\pi GM},
\end{equation}
which must equal the temperature in Eq. \eqref{tempgup}
\begin{equation}
\frac{1+\eta}{8\pi GM}=\frac{1}{8\pi GM}\left(1-\frac{\alpha}{8\pi GM}+5\left(\frac{\alpha}{8\pi GM}\right)^2 \right).
\end{equation}
Solving for $\eta$
\begin{equation}
\eta=-\frac{\alpha}{8\pi GM}+5\left(\frac{\alpha}{8\pi GM}\right)^2,
\end{equation}
and to first order in $\alpha$ the metric is modified by the function
\begin{equation}
\label{Frb}
F(r)=\left(1-\frac{2GM}{r}\right) \left(1-\frac{l_p \alpha_0}{4\pi}\frac{(2GM)^{n-1}}{r^n}\right).
\end{equation}
where we used $\alpha=l_p\alpha_0$.
\par\noindent

A couple of comments are in order here. First, it is easily seen from Eq. (\ref{Frb}) that if one selects the dimensionless constant $\alpha_0$ to be positive (and thus $\eta$ is negative),  then the metric in Eq. (\ref{Frb}) will 
have two horizons as is the case for the modified metric in \cite{Scardigli:2014qka}. However, if one selects the dimensionless constant $\alpha_0$ to be negative (and thus $\eta$ is positive), in this case the metric in Eq. (\ref{Frb}) will have only one  horizon, i.e., $r_s=2GM$.
Second, the value of $n$ is still  undetermined. In the next section, we will determine the value of $n$ from the modified Newton's law.

\section{Modified Newton's Law}
\vspace{-2ex}
\par\noindent
In this section, we calculate the modified Newton's law from the modified metric, 
and we will follow the derivation of gravitational acceleration in (\cite{taylor2000}, p. 3-32).
For a mass falling radially from rest at $r_0$, $d\tau^2=F(r_0)dt^2$; thus, its energy is
\begin{equation}
E=F(r)\frac{dt}{d\tau}=\sqrt{F(r_0)}.
\end{equation}
Substituting $d\tau$ from the previous equation in the metric, Eq. \eqref{metric} with $d\Omega=0$, and solving for $dr/dt$
\begin{equation}
\frac{dr}{dt}=F(r)\sqrt{1-\frac{F(r)}{F(r_0)}}.
\end{equation}
The proper time and length experienced by a static observer on a spherical shell of radius $r$ is given by
\begin{equation}
dt_{sh}=F(r)dt, \qquad dr_{sh}=\frac{dr}{F(r)}.
\end{equation}
Thus,
\begin{equation}
\frac{dr_{sh}}{dt_{sh}}=\sqrt{1-\frac{F(r)}{F(r_0)}}.
\end{equation}
Differentiating with respect to $t_{sh}$ and substituting $r=r_0$ we get the acceleration
\begin{equation}
\label{accel}
g=\frac{{{d}^{2}}{{r}_{sh}}}{d{{t}_{sh}}^{2}}=-\frac{1}{2\sqrt{F(r)}}F'(r_0),
\end{equation}
where $F'(r_0)$ is the derivative of $F(r)$ with respect to $r$ evaluated at $r_0$. Substituting the modified function $F(r)$ and expanding to first order in $\alpha_0$, we get
\begin{align}
\label{modacc}
g=&\frac{GM}{r^2}\left(1-\frac{2GM}{r}\right)^{-1/2} \nonumber\\
&\times\left[1-\frac{(2GM)^{n-2}}{4\pi r^n}\left((1+2n)GM-nr\right)l_p\alpha_0\right]~.
\end{align}
\par\noindent
A couple of comments are in order here. First, equation (\ref{modacc}) reduces to the standard relativistic result 
when $\alpha=0$, and to the Newtonian result after neglecting the relativistic factor $1/\sqrt{1-2GM/r}$. 
Second, the acceleration as given by equation (\ref{modacc})  does not depend on the mass of the falling particle 
which means that the Equivalence Principle is not violated. On the contrary,  Equivalence Principle was violated 
when a different form of GUP was utilized  \cite{Ghosh:2013qra}.
\par\noindent
Thus, the modified Schwarzschild metric leads to the modified Newton's law
\begin{equation}
\label{modNewton}
F_N=\frac{GMm}{r^2}\left[1-\frac{(2GM)^{n-2}}{4\pi r^n}\left((1+2n)GM-nr\right)l_p\alpha_0\right].
\end{equation}

\par\noindent
It should be pointed out here that we have neglected the relativistic factor since we would like to keep the GUP correction terms and the other terms to be up to second order in $GM/r$.

\par\noindent
To estimate the value of $n$, we compare our result with phenomenologically well motivated approaches that modify Newton's law of gravity at short distance
such as Randall-Sundrum II model  \cite{Randall:1999vf} which implies a modification of Newton's law on a brane \cite{Callin:2004py} as follows
\begin{equation}
F_{RS}=
	\begin{cases}
      \frac{GMm}{r^2}\left(1+\frac{4\Lambda_R}{3\pi r}\right), & r\ll\Lambda_R\\ \\
      \frac{GMm}{r^2}\left(1+\frac{2\Lambda_R}{3\pi r^2}\right), & r\gg\Lambda_R
	\end{cases}
\end{equation}
where $\Lambda_R$ is a characteristic length scale.

When $n=2$ in the modified Newton's law of Eq. \eqref{modNewton} we get to first order in $1/r$
\begin{equation}
F_N=\frac{GMm}{r^2}\left(1+\frac{l_p\alpha_0}{2\pi r}\right),
\end{equation}
which clearly agrees with the Randall-Sundrum II result. We conclude that the most likely value for $n$ is 2 and thereon we set $n=2$. Thus, the function $F(R)$ in the modified metric takes the form
\begin{equation}
\label{Fr}
F(r)=\left(1-\frac{2GM}{r}\right) \left(1-\frac{l_p \alpha_0}{2\pi}\frac{GM}{r^2}\right).
\end{equation}
A couple of comments are in order here. First, we obtain these corrections in the framework of semiclassical gravity approach. Thus, we keep the LHS of the Einstein equations as it is and we assign the QG corrections to the RHS of the Einstein equations. Therefore,  using Mathematica and employing the metric element given by Eq. (\ref{Fr}), the RHS of Einstein equations is not zero, as expected. In addition, the RHS of Einstein Equations has only diagonal terms and all of them are perturbation terms of the first order in the perturbative parameter which reads
\begin{equation}
\epsilon = \frac{l_{p} \alpha_{0}}{2\pi}~. \nonumber
\end{equation}
The non-zero diagonal terms  that appear in the RHS of Einstein equations are of the form
\begin{eqnarray}
G^{t}_{t} &=& G^{r}_{r} = + \epsilon \frac{GM (4GM -r)}{r^{5}}\nonumber \\
G^{\theta}_{\theta} &=& G^{\phi}_{\phi} = -\epsilon  \frac{GM (6GM -r)}{r^{5}} \nonumber~.
\end{eqnarray}
Therefore, these  GUP corrections can be  treated as first order perturbation terms around the vacuum solution, i.e., Schwarzschild solution, and the proposed solution can be considered a solution of the Einstein equations  in a perturbative sense. 

\par\noindent
Second, it is evident that the specific expression for  the metric element $F(r)$ implies the existence of another horizon at
\begin{equation}
r=\sqrt{l_p \alpha_0 GM/2\pi},
\end{equation}
which will be an inner horizon. Thus, the spacetime contains an inner and outer horizon, as well as a timelike singularity and the conformal structure should be the same as that of a Reissner-Nordstrom black hole. This implies the existence of a Cauchy horizon which in turn leads to mass inflation.
\section{Deflection of Light}
\vspace{-2ex}
\par\noindent
When light approaches a massive body, such as the sun, it gets deflected from a straight line by an angle given 
by (\cite{weinberg}, p. 189)
\begin{equation}
\label{deflection}
\Delta\phi=2\int_{r_0}^{\infty}\frac{1}{r\sqrt{F(r)}}\left(\frac{r^2}{r_0^2}\frac{F(r_0)}{F(r)}-1 \right)^{-1/2}dr-\pi,
\end{equation}
where $r_0$ is the distance of closest approach to the sun. In general relativity the deflection angle is given 
by (\cite{weinberg}, p. 190)
\begin{equation}
\label{grdefl}
\Delta\phi_{GR}\simeq \frac{4GM}{r_0}.
\end{equation}
\par\noindent
To find the deflection angle predicted by the modified metric, we need to use $F(r)$ from Eq. \eqref{Fr} in Eq. \eqref{deflection}. To simplify the calculations, we make the transformation $u\equiv r_0/r$ in Eq. \eqref{deflection}
\begin{equation}
\label{deflectionu}
\Delta\phi=2\int_{0}^{1}\frac{1}{\sqrt{F\left(\frac{r_0}{u}\right)}}\left(\frac{F(r_0)}{F\left(\frac{r_0}{u}\right)}-u^2 \right)^{-1/2}du-\pi
\end{equation}

To simplify the integral, we expand the integrand to first order in $\alpha_0$ and to second order in $1/r_0$
\begin{align}
\Delta\phi=&\Delta\phi_{GR} \nonumber\\
&+\int_0^2 \frac{u^2 +1}{2\pi r_0 \sqrt{1-u^2}}\frac{GM}{r_0}l_p\alpha_0 du
\end{align}
which evaluates in terms of the gamma function to
\begin{equation}
\Delta\phi=\Delta\phi_{GR}+\frac{\Gamma(\frac{5}{2})}{2\sqrt{\pi}\Gamma(2)}\frac{GM}{r_0^2}l_p\alpha_0.
\end{equation}

\par\noindent
The best accuracy of measuring the deflection of light by the sun is from measuring the deflection of radio waves from distant quasars using the Very Long Baseline Array (VLBA) \cite{Fomalont:2009zg}, which achieved an accuracy of $3\times 10^{-4}$; thus,
\begin{equation}
\frac{\delta\Delta\phi}{\Delta\phi_{GR}}< 3\times 10^{-4}.
\end{equation}
Assuming that light grazes the surface of the sun $r_0\simeq R_\odot=6.96\times 10^8\text{m}=3.53\times10^{24}\text{GeV}^{-1}$ and $M=M_\odot=1.99\times 10^{30}\text{kg}=1.116\times10^{57}\text{GeV}$, we get an upper bound on $\alpha_0$ of
\begin{equation}
\alpha_0< 1.4\times 10^{41}.
\end{equation}

\par\noindent
This bound is larger than the bound set by the electroweak
scale $10^{17}$ but not incompatible with it. However,
studying the effects of the GUP, which is model independent, on
gravitational phenomena might prove useful in understanding the effects
of quantum gravity in that regime.

\section{Time Delay of Light}
\vspace{-2ex}
\par\noindent
In general relativity, the time taken by light to travel from  $r=r_1$ to $r=r_2$ passing by a massive body, 
such as the sun, is slightly longer than what is expected in flat spacetime, and the time of the round trip is given by
\begin{equation}
T=2\left( t(r_1,r_0)+t(r_2,r_0)\right),
\end{equation}
with (\cite{weinberg}, p. 202)
\begin{equation}
\label{delay}
t\left( r,{{r}_{0}} \right)=\int_{r_0}^{r}\frac{1}{F\left( r \right)}{{\left( 1-\frac{F\left( r \right)}{F\left( {{r}_{0}} \right)}\frac{r_0^2}{{{r}^{2}}} \right)}^{-1/2}}\text{d}r,
\end{equation}
and $r_0$ is the distance of closest approach to the sun.

\par\noindent
General Relativity predicts a time of travel (\cite{weinberg}, p. 203)
\begin{align}
T_{GR}=& 2\sqrt{r_1^2-r_0^2}+2\sqrt{r_2^2-r_0^2} \nonumber\\
&+4GM\left(1+\ln\left(\frac{4r_1r_2}{r_0^2}\right) \right)
\end{align}
Using the modified function $F(r)$, we apply the transformation $u\equiv r_0/r$ in Eq. \eqref{delay}
\begin{equation}
t(r_1,r_0)=\int_{\frac{r_0}{r_1}}^{1} \frac{r_0}{u^2F\left(\frac{r_0}{u}\right)}\left(1-u^2 \frac{F\left(\frac{r_0}{u}\right)}{F(r_0)} \right)^{-1/2}du.
\end{equation}
Expanding the integrand to first order in $\alpha_0$ and $M/r_0$ we get
\begin{align}
\Delta T=&\Delta T_{GR}+\left[\int_{\frac{r_0}{r_1}}^{1}\frac{du}{\sqrt{1-u^2}}+\int_{\frac{r_0}{r_2}}^{1}\frac{du}{\sqrt{1-u^2}} \right]\frac{3GMl_p\alpha_0}{2\pi r_0} \nonumber\\
=&\Delta T_{GR}+\left[\pi-\sin^{-1}\frac{r_0}{r_1}-\sin^{-1}\frac{r_0}{r_2}\right]\frac{3GMl_p\alpha_0}{2\pi r_0}.
\end{align}

\par\noindent
The best accuracy of measuring the delay was obtained from the delay in the travel time of radio waves from earth to the Cassini spacecraft \cite{cassini} when it was at a geocentric distance of $8.43\text{AU}=6.39\times10^{27}\text{GeV}^{-1}$, and the closest distance of the photons to the sun was $r_0=1.6R_\odot=5.64\times10^{24}\text{GeV}^{-1}$. The experiment achieved an accuracy of $2.3\times 10^{-5}$, which means that
\begin{equation}
\frac{\delta\Delta T}{\Delta T_{GR}}<2.3\times 10^{-5}.
\end{equation}
setting an upper bound on $\alpha_0$ of
\begin{equation}
\alpha_0 <5.8\times 10^{40}
\end{equation}
which is slightly less than the bound from the deflection of light but still compatible with the bound set by electroweak scale.
%
\section{Perihelion Precession}
\vspace{-2ex}
\par\noindent
In general relativity, the orbit of a particle around a massive body, such as Mercury around the sun, 
precesses in each revolution by an angle given by (\cite{weinberg}, p. 195)
\begin{align}
\label{perihelion}
\Delta\phi&= 2|\phi(r_+)-\phi(r_-)|-2\pi \nonumber \\
&=-2\pi+ 2\int_{r_-}^{r_+}\frac{dr}{r^2\sqrt{F(r)}} \nonumber\\
&\hspace{-1.5ex}\times\left[\frac{r_-^2\left(1 -\frac{F(r)}{F(r_-)}\right)-r_+^2\left(1 -\frac{F(r)}{F(r_+)}\right)}{r_+^2r_-^2\left(\frac{F(r)}{F(r_+)} -\frac{F(r)}{F(r_-)}\right)} -\frac{1}{r^2}\right]^{-\frac{1}{2}}
\end{align}
where $r_-$ and $r_+$ are the minimum and maximum values of $r$ respectively. 
The precession predicted by general relativity is given by (\cite{weinberg}, p. 197)
\begin{equation}
\label{periGR}
\Delta\phi_{GR}\simeq 3\pi GM\left(\frac{1}{r_+}+\frac{1}{r_-} \right).
\end{equation}

\par\noindent
Using the modified function $F(r)$, we change the integration variable in Eq. \eqref{perihelion} to $u\equiv r_-/r$
\begin{align}
\Delta&\phi=-2\pi+2\int_{r_-/r_+}^{1}\frac{du}{{{r}_{-}}\sqrt{F\left(\frac{r_-}{u}\right)}} \nonumber\\
&\times\left[ \frac{r_{-}^{2}\left( 1-\frac{F\left(\frac{r_-}{u}\right)}{F(r_-)} \right)-r_{+}^{2}
\left( 1-\frac{F\left(\frac{r_-}{u}\right)}{F(r_+)} \right)}{r_{+}^{2}r_{-}^{2}\left( \frac{F\left(\frac{r_-}{u}\right)}
{F(r_+)}-\frac{F\left(\frac{r_-}{u}\right)}{F(r_-)} \right)}-\frac{{{u}^{2}}}{r_{-}^{2}} \right]^{-\frac{1}{2}}.
\end{align}
\par\noindent
We expand the integrand to first order in $\alpha$ and to first order in $(1/r_{+}  + 1/r_{-})$ to obtain
\begin{align}
\Delta\phi&=\Delta\phi_{GR}+\frac{1}{4\pi}\int_{r_-/r_+}^{1}\frac{\left(r_+^{-1}+r_-^{-1}\right)l_p\alpha_0}{\sqrt{(u-1)\left(\frac{r_-}{r_+}-u\right)}} du\nonumber \\
&=\Delta\phi_{GR}+\frac{1}{4}\left(\frac{1}{r_+}+\frac{1}{r_-} \right)l_p\alpha_0
\end{align}

\par\noindent
The best measurement of the perihelion precession of Mercury is $42.980\pm 0.002$ arcsec per century \cite{Pireaux:2001yk,Will:2014xja}, which amounts to an accuracy of $4.6\times 10^{-5}$. Thus, we have
\begin{equation}
\frac{\delta\Delta\phi}{\Delta\phi_{GR}}<4.6\times 10^{-5}.
\end{equation}
Using the perihelion and aphelion of Mercury $r_-=4.60\times10^{10}\text{m}=2.33\times10^{26}\text{GeV}^{-1}$ and $r_+=6.98\times10^{10}\text{m}=3.54\times10^{26}\text{GeV}^{-1}$, we get an upper bound on $\alpha_0$ of
\begin{equation}
\alpha_0<1.6\times 10^{35}
\end{equation}
which is five orders of magnitude less than the bound from the time delay and the deflection of light.

\section{Gravitational Redshift}
\vspace{-2ex}
\par\noindent
Suppose that light was emitted from radius $r_1$ and received at $r_2$; by how much will the light be red-shifted? In the metric \eqref{metric} put $dr=0$ and $d\phi=0$, and solve for $dt$. Since the time measured by a remote observer is the same for the two radii, we get
\begin{equation}
dt=\frac{d\tau_1}{\sqrt{F(r_1)}}=\frac{d\tau_2}{\sqrt{F(r_2)}},
\end{equation}
and the relative frequency is
\begin{equation}
\frac{\omega_2}{\omega _1}=\frac{d\tau_1}{d\tau _2}=\sqrt{\frac{F(r_1)}{F(r_2)}}.
\end{equation}
Substituting $F(r)$ from \eqref{Fr}, and expanding to first order in $\alpha_0$
\begin{align}
\frac{\omega_2}{\omega _1}=&\sqrt{\frac{1-2GM/r_1}{1-2GM/r_2}}\nonumber \\
&\left(1+ \frac{GM}{4\pi }\left[\frac{1}{r_2^2}-\frac{1}{r_1^2} \right]l_p\alpha_0 \right).
\end{align}
Subtracting one from the previous result we get
\begin{align}
\label{redshift}
\frac{\omega_2-\omega_1}{\omega_1}=&(S-1)\nonumber\\
&\left(1+\frac{S}{S-1}\frac{GM}{4\pi}\left[\frac{1}{r_2^2}-\frac{1}{r_1^2} \right]l_p\alpha_0\right),
\end{align}
where
\begin{equation}
S\equiv\sqrt{\frac{1-2GM/r_1}{1-2GM/r_2}}.
\end{equation}

\par\noindent
The most accurate measurement of gravitational redshift is from Gravity Probe A \cite{Vessot:1980zz} in 1976. The satellite was at an altitude of $10^7 \text{m}=5.07\times10^{22}\text{GeV}^{-1}$, and achieved an accuracy of $7.0\times 10^{-5}$, which means that the new term in Eq. \eqref{redshift}
\begin{equation}
\frac{S}{S-1}\frac{GM}{4\pi}\left[\frac{1}{r_2^2}-\frac{1}{r_1^2} \right]l_p\alpha_0<7.0\times 10^{-5}.
\end{equation}
Using the mass and radius of the earth, $M_\oplus=5.97\times10^{24}\text{kg}=3.35\times10^{51}\text{GeV}$, $ r_1=R_\oplus=6.38\times10^{6}\text{m}=3.23\times10^{22}\text{GeV}^{-1}$, and $r_2=r_1+10^7 \text{m}$. The corrections are negative since $r_2>r_1$ and hence the bound is given as follows
\begin{equation}
\alpha_0<2.5\times 10^{38}.
\end{equation}
This bound is stringent too and compatible with the bound set by the electroweak scale.

\begin{table}[h]
\caption{Bounds on the GUP parameter $\alpha_0$ from gravitational tests}
\label{Table1}
\begin{ruledtabular}
\begin{tabular}{lc}
Experiment & Bound on $\alpha_0$ \\
\hline Deflection of Light& $1.4\times 10^{41}$ \\  Time Delay of Light & $5.8\times 10^{40}$ \\
Perihelion Precession& $1.6\times 10^{35}$ \\
Gravitational Redshift& $2.5\times 10^{38}$ \\
\end{tabular}
\end{ruledtabular}
\end{table}

\par\noindent
It is noteworthy that comparing the version of GUP we employ in our analysis here (see Eq. (\ref{gupsimple})) and the corresponding one used in \cite{Scardigli:2014qka} (see Eq. (17) in  \cite{Scardigli:2014qka}), one can say that the corresponding GUP parameters, i.e., $\alpha_{0}$ and $\beta$, respectively, are similar, namely $\alpha_{0}^{2}\sim \beta$. Utilizing this similarity,  one can conclude that the bounds on the two aforesaid GUP parameters based on the tests for the deflection of light and the perihelion precession are essentially equivalent. 
\section{Conclusions}
\vspace{-2ex}
\par\noindent
In this paper, we proposed a modification to the Schwarzschild metric based on the GUP in Eq. (\ref{gup}) 
to reproduce the modified Hawking temperature derived from the GUP. This modification preserves 
the horizon at $2GM$, and predicts the existence of another horizon which might be the radius of the black hole singularity. 
We assumed a modification with a general $1/r^n$ dependence, and determined the value $n=2$ by comparing the modified 
Newton's law derived from the modified metric with phenomenologically well-motivated approaches that modify 
Newton's law of gravitation at short distance such as the Randall-Sundrum II model. We computed corrections 
to the general relativistic results of the deflection of light, time delay of light, perihelion precession, 
and gravitational redshift. We compared our results with measurments to obtain upper bounds on the GUP 
parameter $\alpha_0$ (see Table \ref{Table1}). The bounds we found in this paper are greater than those 
reported in previous work from corrections to quantum mechanical predictions 
\cite{Ali:2009zq,Das:2010zf,Ali:2011fa,Das:2011tq}. However, investigating the implications of the 
GUP on gravitational phenomena might prove useful for understanding the effects of quantum gravity in that regime. 
In addition, because the GUP is model independent, this understanding can help to evaluate the results of different theories 
of quantum gravity.\\
%
\section{Acknowledgments}
\par\noindent
The authors would like to thank the referee for the constructive comments.



\end{document}